**Narrow genetic base in forest restoration with holm oak (*Quercus ilex* L.) in Sicily**

Burgarella Concetta[1*], Navascués Miguel[2,3], Soto Álvaro[4,5], Lora Ángel[6], Fici Silvio[1]

[1]Dip.to di Scienze Botaniche, Universitá di Palermo, Via Archirafi 38, 90123 Palermo, Italy

[2]School of Biological Sciences, University of East Anglia, Norwich NR4 7TJ, UK

[3]Current address: Laboratoire d'Ecologie, École Normale Supérieure, 46 Rue d'Ulm 75230 Paris, France

[4]Dpto. Sistemas y Recursos Forestales, CIFOR-INIA, P.O. Box 8111, 28080, Madrid, Spain

[5]Current address: Dpto. Silvopascicultura, ETSI Montes, Universidad Politécnica de Madrid. Ciudad Universitaria s/n 28040 Madrid, Spain

[6]Dpto. de Ingeniería Forestal, ETSIAM, Universidad de Córdoba, Av. Menéndez Pidal s/n, 14080 Córdoba, Spain

* Corresponding author

Phone number: +390916238205

Fax: +390916238203

E-mail: cr1burco@uco.es

**Running title**: Genetic diversity in holm oak restoration

**Abstract**

In order to empirically assess the effect of actual seed sampling strategy on genetic diversity of holm oak (*Quercus ilex*) forestations in Sicily, we have analysed the genetic composition of two seedling lots (nursery stock and plantation) and their known natural seed origin stand by means of six nuclear microsatellite loci. Significant reduction in genetic diversity and significant difference in genetic composition of the seedling lots compared to the seed origin stand were detected. The female and the total effective number of parents were quantified by means of maternity assignment of seedlings and temporal changes in allele frequencies. Extremely low effective maternity numbers were estimated ($Nf_e \approx 2–4$) and estimates accounting for both seed and pollen donors give also low values ($N_e \approx 35–50$). These values can be explained by an inappropriate forestry seed harvest strategy limited to a small number of spatially close trees.

*Quercus ilex* **/ plantation / genetic diversity / effective population size / microsatellite**

**Introduction**

The maintenance of the natural patterns of genetic diversity of population and species has been widely recognized as a key factor for the preservation of their evolutionary potential [30]. The use of autochthonous material is recommended for common forest practices [24], but there are no guidelines on how much genetic diversity in natural populations should be represented in an artificially reforested stand to guarantee its viability in the following generations [30]. The introduction of low diversity material could result in a reduced long term viability of plantations or in the failure of demographic rescue of local impoverished populations, due to a decrease in their effective population size [25].

Theoretical approaches establish that the genetic drift in the seed collection process is determined by the number of seed parents before than by the number of seeds per parent [4, 43]. In wild seed collection the effective size and diversity of pollen donors is unknown a priori, hence the number of seed trees definitively represents the operative tool for achieving the conservation of genetic diversity levels [13]. Studies exploring the genetic diversity of plantations originated from seedlots collected from natural stands are scarce. In some cases, reduction or biases in genetic composition of plantations have been linked to a limited or non-random sampling of maternal trees [13, 26, 35, 40], although this assumption was not verified experimentally.

On Sicily, natural woods currently occupy only 10% of the area and one third of them are broadleaf formations. At present, Sicilian woods are scarcely economically productive, their main interest lying on ecological conservation and landscape values. Mediterranean vegetation dominates Sicilian ecosystems, where the holm oak, *Quercus ilex* L., is a key species in many primary and secondary formations from sea level up to 1 800 m. On the island, this schlerophyllous evergreen tree forms pure and mixed forests, although it is locally reduced to small relict populations. Over its western Mediterranean distribution the holm oak shows high levels of nuclear genetic diversity within populations, and low interpopulation differentiation [31]. Maternally inherited chloroplast genome suggested glacial refugia in the three Mediterranean peninsulas and Sicilian populations have shown to represent a reservoir of diversity [11].

In order to empirically assess whether forest genetic material from actual seed sampling strategies suffer changes in genetic diversity relative to natural old-growth populations, we analysed the genetic composition of seedling lots in comparison with the known autochthonous seed origin stand. We used *Quercus ilex* as model tree. Its representative role in Sicilian natural and artificial forest ecosystems makes it one of the most widely used species for restoring deforested areas and converting introduced pine plantations. By testing intrapopulation genetic diversity measures and quantifying genetic drift effects we discuss how actual forestry practices could affect the long term viability of holm oak plantations in Sicily.

**Materials and methods**

*Study site and experimental design.*

The study was undertaken in Sicily (Fig. 1): one-year-old seedlings from a plantation located in the Monte Palmeto and from the nursery of Piano Noce were sampled. These two progeny sets were originated from seeds collected in 2001 on the ground of a 100 m × 60 m enclosure containing 15 adult holm oaks (hereafter called candidate mother trees). This group of trees is used as seed source for Piano Noce nursery, which supplies the holm oak seedling demand for reforestation actions in the north-western part of Sicily. The 15 individuals and 40 additional adult trees from the surrounding continuous natural forest of Piano Zucchi were collected (sample referred as seed origin stand from now on, trees outside the enclosure were chosen randomly maintaining 50 m distance between them).

[[[Figure 1]]]

Sample size for the progeny sets were set to 40 individuals, but only 33 individuals were available from Monte Palmeto because of the high mortality rate for seedlings after plantation. Fresh leaves were collected from each individual and stored at -80ºC. DNA was extracted following the method described by Doyle and Doyle [7].

*Molecular markers*

All individuals of the study were genotyped for six microsatellite loci: MSQ4, developed for *Quercus macrocarpa* [5]; QpZAG15, QpZAG36 and QpZAG46 developed for *Q. petraea* [39]; QrZAG11 and QrZAG20 developed for *Q. robur* [20]. Amplification was performed as described in SOTO *et al.* [37], except for QpZAG36 and QrZAG20. The annealing temperature of 51ºC has been used with QpZAG36. A touchdown procedure has been used for QrZAG20, consisting in 20 cycles starting at 65ºC and decreasing 0.5ºC each cycle, followed by 20 cycles at 55ºC. PCR products were sized in 6% polyacrylammide gels and electrophoresis was performed on an automatic sequencer Li-Cor 4200 (Li-Cor Biosciences). Microsatellites were scored with Gene ImagIR v. 3.56 (Li-Cor Biosciences).

*Assessment of Hardy-Weinberg model*

Preliminary analysis were conducted with MICRO-CHECKER 2.2.0 [42] to assess the possibility of null alleles or genotyping errors due to stuttering or allelic drop out. Every pair of loci was tested for linkage disequilibrium by using FSTAT 2.9.3.2 [14], because the independent transmission of alleles is a required condition for subsequent analyses (estimation of kinship and genetic differentiation, parentage analysis). Two of our microsatellites loci belong to the same linkage group in *Quercus robur* [3], but their linkage in *Q. ilex* has never been studied. According to the results of linkage disequilibrium test, null allele estimation and exclusion probability computation, we decided to exclude locus QpZAG36 from analyses requiring unlinked loci (see

130 results). Single locus genotypes were tested for deviations from Hardy–Weinberg

131 expectations by using FSTAT 2.9.3.2 (1 000 permutations), to assess whether

132 inbreeding or familiar relationships might produce interferences with the linkage

133 disequilibrium analysis. The fixation index $F_{IS}$ was calculated for each locus and

134 overall loci with the same program. Since the presence of a familiar structure inside

135 the 15-holm-oak enclosure of Piano Zucchi could produce a reduced genotype

136 variability in the offspring, as well as some bias in parentage assignment [29], the

137 relationship among the candidate mother trees has been examined by the estimation

138 of the kinship coefficient, $F$ [27], with SPAGEDI 1.2 [16].

139

140 *Genetic diversity and differentiation*

141 The following indices were computed for each locus and for each sample: number of

142 alleles, $n_a$; allelic richness standardized to the smallest sample, $A$ [9]; unbiased

143 effective number of alleles, $A_e$ [34] and unbiased gene diversity, $H_e$ [33]. To explore

144 whether diversity indices in the seedling lots had lower values compared to the

145 natural seed origin stand, a Monte Carlo resampling approach has been used (10 000

146 iterations) for each pair of compared samples (pair 1: seed origin stand/nursery stock;

147 pair 2: seed origin stand/plantation). This approach provided an estimation of the p-

148 value to reject the null hypothesis of no difference in genetic diversity levels among

149 samples. Whole multilocus genotypes were permuted to maintain the original

150 association of allele within the genotypes. The pairwise genetic differentiation $\theta$ [45]

and its significance have been estimated with a permutation procedure (10 000 iterations) with FSTAT 2.9.3.2 software, assuming Hardy-Weinberg equilibrium.

*Parentage inference and effective number of mothers*

The power of our set of loci in parentage assignments was evaluated computing the value of non-exclusion probabilities [18] for one parent when the genotype of the other parent is unknown and for parent pair with CERVUS 3.0 [19]. In order to infer the number of individuals, among the 15 candidate mothers, contributing to the genetic diversity of the two seedling sets, two types of parentage analysis was undertaken with CERVUS. The program uses a likelihood-based approach in which the strength of parentage assignment is evaluated with the log-likelihood ratio calculated over all loci (LOD-score) for each candidate parent. Using a simulation procedure CERVUS produces a critical LOD-score value, below which parentage cannot be attributed at the level of precision chosen (here 80% and 95% were used). A value of 0.001 has been used for the error rate to take into account the occurrence of mistyping, null alleles or mutations. Allele frequencies from seed origin stand were used as reference for CERVUS calculations. We first performed a one parent analysis, where CERVUS searches for the first parent in absence of genetic information on the second parent. We assumed the most likely parent assigned being the mother in consideration of two aspects: i) the knowledge that both seedling lots proceeded from the 15-trees enclosure and ii) the low probability of finding the fathers inside the enclosure, taking into account its restricted area and the high level of pollen flow expected in *Quercus*. The sampled percentage of breeding female population was set to 95%. We considered 95% a conservative estimation as we cannot exclude the dispersal of seeds from outside the 15-tree enclosure. In any case, a 100% value was also used to assess the effect of this parameter on the percentage of unresolved assignments. With the same set of 15 candidate parents a parent pair analysis was also carried out, in order to quantify, if any, the bias in maternity assignments due to the identification of male parents among the first-parent assignments. We set the 15 trees both as candidate male parents and candidate female parents. As a conservative approach, the proportion of female candidate parents sampled was set to 0.95, while proportion of male candidate parents sampled was set to 0.50. An independent control of parentage assignment was made taking advantage of the putative linkage between loci QpZAG46 and QpZAG36, comparing the match of association between alleles from each individual of the offspring and the inferred mother tree or parent pair.

The reproductive success per mother tree, estimated with maternity analysis with more than 80% of confidence, allowed to compute the effective maternity number, $Nf_e$ [34]. Furthermore, to evaluate the potential resistance to random genetic drift of seedling lots included in the nursery stock of Piano Noce and involved in the plantation of Monte Palmeto, effective population sizes ($N_e$) for the two seedling samples have been estimated with a likelihood procedure implemented in MLNE 1.0 [44], which takes into account the changes in allelic frequencies between two generations (considering the seed origin stand, Piano Zucchi, as the first generation).

**Results**

*Assessment of Hardy-Weinberg model*

Only one locus (QpZAG36) for one sample (the seed origin stand, Piano Zucchi) showed a significant excess of homozygotes (p = 0.001), which could be due to the presence of null alleles, however at least one allele was amplified for all individuals at this locus. No evidence of stuttering or allelic drop out for larger alleles could be detected. For all samples test for Hardy–Weinberg equilibrium revealed a significant departure (p = 0.002) only for QpZAG36 among the six loci used. No evidence of inbreeding has been found (data not shown). Close relatedness among the 15 candidate seed mother trees was not detected, as their mean kinship value ($\overline{F_{ij}}$ = -0.003) is similar to the mean value relative to all the other individuals sampled in the whole seed origin stand ($\overline{F_{ij}}$ = -0.008). Linkage disequilibrium was significant between loci QpZAG46 and QpZAG36 in all samples (p = 0.008). We will consider likely that these loci maintain some level of genetic linkage in *Q. ilex* genome.

*Genetic diversity and differentiation*

The indices tested show that genetic diversity in both seedling lots has suffered a significant reduction compared to natural seed origin stand (Table I). In particular, allelic richness decreases by 22.5% and 33.6%, and the effective number of alleles by 18.9% and 32.5%, in the nursery and plantation respectively. The expected heterozygosity ($H_e$) shows a similar pattern of reduction, but less dramatic. In fact, only the reduction suffered by the planted stand results significant. The artificial populations showed also a significantly different genetic composition compared with the seed origin stand, although the significance of the pairwise genetic differentiation, $\theta$ (Table I), might be overestimating the differentiation, due to the particular set of markers used [as shown in 36].

[[[Table 1]]]

*Maternity inference and effective number of mothers*

One parent assignment was performed by a maternity analysis to infer the number of trees among the 15 candidate mothers that had contributed to the genetic diversity observed in the nursery sample and in the plantation. The exclusion of QpZAG36 locus in the maternity analysis because of its putative linkage to QpZAG46 does not substantially change the parentage inference in the light of the combined non-exclusion probabilities estimated (Table II).

[[[Table 2]]]

At 80% of confidence 98% of the seedlings could be ascribed to one of the candidate mothers (97% of plantation individuals and 100% individuals from the nursery sample). Within the assigned offspring over 17% matched at 95% of confidence. For the 0.001 error rate assumed, only one individual of the plantation are left unassigned, setting the sampled percentage of the breeding female population either to 95% or to 100%. Among the 15 candidate mothers seven have been identified as source of seeds for the nursery sample and eight for the plantation (Figure 2). Most of the seedlings sampled (58%) were identified as offspring of mother trees m1, m9, m12. However, assigned mothers are not shared between the two offspring groups. In fact, within the nursery sample 78% individuals come from trees m1 and m9, while within the plantation 32% come from trees m12 and 43% from trees m14 and m15 (Figure 2). The parent pair assignment identifies only one parental pair for one seedling (among 66 analyzed) at 80% of confidence. This does not lead to any correction of the maternity assignment, as the two genotypes correspond to the same candidate parent (m1).

[[[Figure 2 ]]]

Almost every pair offspring-mother genotypes have been confirmed by the control of the match of allele association for the putative linked loci QpZAG46 and QpZAG36. For all but six seedlings we observed an allele combination concordant with the assigned mother genotype, and, when more than one seedling was attributed to the same candidate mother, this was true for the whole group of seedlings (data not shown). The six mismatches found could be due to either the error rate included in the mother assignment method or the incorrect identification of alleles during genotyping. However, among the six mismatches, one individual from the nursery sample and three from the planted stand match with the second most likely mother tree. In any case, considering the second most likely mothers as the true ones for these seedlings would not increase the total number of assigned mothers. The match of allele association was also compatible for the only trio offspring-parent pair genotypes.

Maternity analysis results expressed in terms of the female contribution to parental population produced very low estimations of effective maternity number ($Nf_e$). For the nursery sample a value of $Nf_e = 1.65$ corresponds to the seven putative mother trees, while for the plantation a value of $Nf_e = 4.39$ corresponds to the eight putative mothers. The effective population size accounting for both female and male contribution gives $N_e = 52.5$ (95% confidence interval, CI, 32.1–120.0) for the nursery sample and $N_e = 35.4$ (95% CI 24.3–61.7) for the plantation.

**Discussion**

*From natural seed origin stand to plantations*

Ecosystem restoration is an issue of major concern to forest management in Sicily and, regarding *Quercus ilex* genetic resources, priority has to be given to a conservation forestry to preserve the specific genetic diversity found in Sicilian populations [11, 28, 31]. Quite the contrary, the results of this study showed an overall significant reduction of genetic diversity and a significant difference in genetic composition of seedling lots in comparison with the natural seed origin stand. The complementary approach of maternal assignment, may be more informative for the purposes of this study because it allows direct measure of the reduction in population size, revealed that very few mother trees have finally contributed to the genetic diversity of artificial populations examined (seven trees for the nursery stock and eight for the plantation). In terms of effective number of seed donors, the number of contributing mother trees is further reduced because of the differences in reproductive success. Estimates accounting for both seed and pollen donors show also low values. Comparing the genetic behaviour of experimental populations when founder events were induced [10, 38] with our finding of reduced population effective size in holm oak seedling lots, we deduce that the latter experienced the equivalent to a bottleneck process, due primarily to collecting seeds from a limited number of mother trees.

Our results conform to previous works which have found that seedlots or planted stands have a reduced genetic diversity or different genetic composition in comparison with natural unmanaged populations. In some cases seed harvest from few or non-randomly selected trees has been proposed as one of the most likely causes [13, 26, 35]. The importance of small population size effect was also shown by Kitzmiller [21], who compared seedlots of *Pinus ponderosa* and reported the lowest allelic diversity for the lot collected from the smallest number of trees.

*Factors involved in genetic composition of seedlings*

Practical considerations usually determine the selection of seed harvest site in Sicily, including nursery proximity and road accessibility. Currently, no genetic criteria are taken into consideration, either because of the general lack of knowledge on the level and distribution of genetic diversity in Sicilian forest species, or because of the absence of legal regulations of the genetic composition of seedlots. In theory, according to *Quercus* species features (allogamy and wind pollination) 15 trees could be a suitable number to collect the great majority of the population variability [4]. Nevertheless, the strong difference between the number of candidate mother trees and the actual contributing mothers indicate that other natural and artificial factors, not taken into consideration for collection planning, could have had a significant contribution in reducing seedlots diversity.

In general, reproductive properties as asynchrony in flowering and fruiting phenology among plants, and individual inter-annual variation in fertility for *Quercus* species [2, 8, 17, 22, 23] determine non-random mating in each reproductive season. In *Quercus ilex*, Lumaret *et al.* [28] recorded that, in a single year, variation in male and female investment involved 15–20% of individuals. In addition, at local scale, differentiation among the pollen clouds received by different mother trees could significantly depend on intermate distance, regardless of its dependence on long distance pollen dispersal [6, 22, 32, 41]. In our study area, the 15-holm-oak enclosure is included in a wider zone characterized by a low density formation which progressively turns into a closed wood. Therefore, likely few closer individuals would have the highest probability of a successful mating in a single reproductive event. Further, the 15 oaks can be classified into two cohorts, 10 very old trees, and five young trees (Figure 1). Since fecundity and acorn production are positively correlated with plant or crown size [1, 15], individuals are expected to differ greatly in their contribution to the next generation in both male and female fertility. Additionally, the overlapping of crowns of some old individuals (m3, m4, m5) among them and with an equal-size maple (*Acer monspessulanum*), may restrict flower and fruit development due to space competition or light limitation [1, 22]. In fact, it is remarkable that the large tree m5, the closer to the maple (Figure 1), does not contribute to any of the seedling groups (Figure 2). All cited factors affecting the individual reproductive performance could have produced a natural bias in the genetic composition of the 15-trees annual acorn production and, therefore, in harvested seedlots.

After collection, genetic variation might be further reduced due to seed and seedling handling and to plant responses to domestication [21, 24, 40] until successful seedling in-field establishment is accomplished. More important might be the maladaptation to

346  local conditions of transplanted material [25] (e.g. altitude difference between the

347  seed origin stand and the plantation is about 700 m). Nevertheless, our data concern

348  neutral genetic diversity, thus the impact of selection cannot be estimated. We have

349  no current data on mortality rate for holm oak seedlings in this study, but it is relevant

350  that Monte Palmeto plantation sample size was constrained to the first-year-survivor

351  seedlings (33 plants over 1 000 initially planted). The post-nursery selection, whether

352  human or environment mediated, could have also led to the shift of coincidence in

353  assigned mother trees between the plantation and the nursery sample.

354

355  *Management implications*

356  The low genetic diversity found for seedlots in this study is likely to concern many

357  recent forestations on Sicily. In the case of *Q. ilex* acorn harvesting from Piano

358  Zucchi forest, an increase in the number of seed trees and distance between trees is

359  recommended. In consideration of the wide extension of Piano Zucchi forest (more

360  than 1 000 hectares), probably the most effective harvest design includes at least 20–

361  30 scattered plants, distributed in a few high distance groups (hundred of meters) of

362  low distance trees (tens of meters). The most efficient model for seed collection and

363  sampling optimization (i.e. minimal number of tree and seeds per tree for the

364  maximal yield) could be reached comparing the seedlot genetic diversity from a

365  number of seed trees progressively higher. In order to achieve this target, setting

366  minimum species-specific levels of diversity for plantations has been devised as a

367  difficult key task [12, 21] since it is subjected to the knowledge of the genetic structure of natural stands in an area which is not available in general (except for few well known temperate species). The genetic diversity of the autochthonous seed origin stand could be the natural baseline for any plantation, as shown by this study [but see 40]. It is straightforward that the ideal situation would also be able to ensure adaptation of genetic material to plantation site [25], but this kind of information seems to be even more difficult to obtain.

Our results are based on a single-year seed collection. However, differences among years are expected as discussed above. In multi-year restoration projects, the annual addition of seedlings could reduce the loss of variability, increasing progressively the effective population size and the genetic base (this could be the case of Monte Palmeto plantation, whose planted area is increased annually). Nevertheless, if plantation is carried out with only a one-season seed stock, its reduced genetic diversity could compromise or make ineffective the restoration aim in the long run (i.e. in isolated condition, in genetic rescuing actions, or under hard environmental conditions).

385 **Acknowledgments**

386 We wish to thank R. Alía for valuable suggestions and discussion on the project. We

387 are also grateful to the Plant Genetics Laboratory of the Centre for Agricultural

388 Formation and Research of Córdoba (IFAPA) for sharing laboratory resources and to

389 Z. Lorenzo for her valuable help on the laboratory work. We also thank two

390 anonymous reviewers for useful suggestions improving the manuscript. The

391 European Social Fund provided a PhD scholarship to C.B.

392
19

REFERENCES

[1] Abrahamson W.G., Layne J.N., Relation of ramet size to acorn production in five oak species of xeric upland habitats in south-central Florida, Am. J. Bot. 89 (2002) 124-131.

[2] Bacilieri R., Ducousso A., Kremer A., Genetic, morphological, ecological and phenological differentiation between *Quercus petraea* (Matt.) Liebl. and *Quercus robur* L. in a mixed stand of northwest of France, Silvae Genet. 44 (1995) 1-10.

[3] Barreneche T., Bodénès C., Lexer C., Trontin J.F., Fluch S., Streiff R., Plomion C., Roussel G., Steinkellner H., Burg K., Favre J.M., Glössl J., Kremer A., A genetic linkage map of *Quercus robur* L. (pedunculate oak) with RAPD, SCAR, microsatellite, minisatellite, isozyme and rDNA markers, Theor. Appl. Genet. 97 (1998) 1090-1103.

[4] Brown H.D., Hardner C.M., Sampling the gene pools of forest trees for ex situ conservation, in: Brown H.D., Hardner C.M. (Eds.), Forest conservation genetics. Principles and practice, CABI Publishing, Wallingford. Oxon, U.K., 2000, pp. 185 - 196.

[5] Dow B., Ashley M., Howe H., Characterization of highly variable (GA/CT)n microsatellites in the bur oak, *Quercus macrocarpa*, Theor. Appl. Genet. 91 (1995) 137-141.

[6] Dow B., Ashley M., High levels of gene flow in bur oak revealed by paternity analysis using microsatellites, J. Hered. 89 (1998) 62-70.

415 [7] Doyle J.J., Doyle L.J., A rapid DNA isolation procedure for small quantities of

416 fresh leaf tissue, Phytoch. Bull. 19 (1987) 11-15.

417 [8] Ducousso A., Michaud H., Lumaret R., Reproduction and gene flow in the genus

418 *Quercus* L., Ann. Sci. For. 50 (1993) 91-106.

419 [9] El Mousadik A., Petit R., High level of genetic differentiation for allelic richness

420 among populations of the argan tree [*Argania spinosa* (L.) Skeels] endemic to

421 Morocco, Theor. Appl. Genet. 92 (1996) 832-839.

422 [10] England P.R., Osler G.H.R., Woodworth L.M., Montgomery M.E., Briscoe

423 D.A., Frankham R., Effects of intense versus diffuse population bottlenecks on

424 microsatellites genetic diversity and evolutionary potential, Conserv. Genet. 4 (2003)

425 595-604.

426 [11] Fineschi S., Cozzolino F., Migliaccio M., Musacchio A., Innocenti M.,

427 Vendramin G.G., Sicily represents the Italian reservoir of chloroplast DNA diversity

428 of *Quercus ilex* L. (Fagaceae), Ann. For. Sci. 62 (2005) 79-84.

429 [12] Finkeldey R., Ziehe M., Genetic implications of silvicultural regimes, For. Ecol.

430 Manag. 197 (2004) 231-244.

431 [13] Gömöry D., Effects of stand origin on the genetic diversity of norway spruce

432 (*Picea abies* Karst.) populations, For. Ecol. Manag. 54 (1992) 215-223.

433 [14] Goudet J., FSTAT, a program to estimate and test gene diversities and fixation

434 indices (version 2.9.3), (2001)

435 [15] Greenberg C.H., Individual variation in acorn production by five species of

436 southern Appalachian oaks, For. Ecol. Manag. 132 (2000) 1999-210.

[16] Hardy O.J., Vekemans X., SPAGeDi: a versatile computer program to analyse spatial genetic structure at the individual or population levels, Mol. Ecol. Notes 2 (2002) 618-620.

[17] Healy W.M., Lewis A.M., Boose E.F., Variation of red oak acorn production, For. Ecol. Manag. 116 (1999) 1-11.

[18] Jamieson A., Taylor S.C.S., Comparison of the tree probability formulae for parentage exclusion, Anim. Gen. 28 (1997) 397-400.

[19] Kalinowski S.T., Taper M.L., Marshall T.C., Revising how the computer program CERVUS accommodates genotyping error increases success in paternity assignment, Mol. Ecol. 16 (2007) 1099-1106.

[20] Kampfer S., Lexer C., Glössl J., Steinkellner H., Characterization of (GA)n microsatellite loci from *Quercus robur*, Hereditas 129 (1998) 183-186.

[21] Kitzmiller J.H., Managing genetic diversity in a tree improvement program, For. Ecol. Manag. 35 (1990) 131-149.

[22] Knapp E.E., Goedde M.A., Rice K.J., Pollen-limited reproduction in blue oak: implications for wind pollination in fragmented populations, Oecologia 128 (2001) 48-55.

[23] Koenig A.O., Mumme R.L., Carmen W.J., Stanback M.T., Acorn production by oaks in central coastal California: variation within and among years, Ecology 75 (1994) 99-109.

[24] Ledig F.T., Human impacts on genetic diversity in forest ecosystems, Oikos 63 (1992) 87-108.

[25] Lefèvre F., Human impacts on forest genetic resources in the temperate zone: an updated review, For. Ecol. Manag. 197 (2004) 257-271.

[26] Li Y.Y., Chen X.Y., Zhang X., Wu T.Y., Lu H.P., Cai Y.W., Genetic differences between wild and artificial populations of *Metasequoia glyptostroboides*: implications for species recovery, Conserv. Biol. 19 (2005) 224-231.

[27] Loiselle B.A., Sork V.L., Nason J.D., Graham C., Spatial genetic structure of a tropical understory shrub, *Psychotria officinalis* (Rubiaceae), Am. J. Bot. 82 (1995) 1420-1425.

[28] Lumaret R., Yacine A., Berrod A., Romane F., Xian Li T., Mating system and genetic diversity in holm oak (*Quercus ilex* L. Fagaceae), in: Lumaret R., Yacine A., Berrod A., Romane F., Xian Li T. (Eds.), Biochemical markers in the population genetics of forest trees, SPB Academic Publishing bv, The Ague, The Netherlands, 1991, pp. 149-153.

[29] Marshall T.C., Slate J., Kruuk L.E.B., Pemberton J.M., Statistical confidence for likelihood-based paternity inference in natural populations, Mol. Ecol. 7 (1998) 639-655.

[30] McKay J.H., Christian C.E., Harrison S., Rice K.J., "How local is local?" A review of practical and conceptual issues in the genetics of restoration, Restor. Ecol. 13 (2005) 432-440.

[31] Michaud H., Toumi L., Lumaret R., Li T.X., Romane F., Di Giusto F., Effect of geographical discontinuity on genetic variation in *Quercus ilex* L. (holm oak). Evidence from enzyme polymorphism, Heredity 74 (1995) 590-606.

[32] Nakanishi A., Tomaru N., Yoshimaru H., Manabe T., Yamamoto S., Interannual genetic heterogeneity of pollen pools accepted by *Quercus salicina* individuals, Mol. Ecol. 14 (2005) 4469-4478.

[33] Nei M., Estimation of average heterozygosity and genetic distance from a small number of individuals, Genetics 89 (1978) 583--590.

[34] Nielsen R., Tarpy D.R., Reeve K., Estimating effective paternity number in social insects and the effective number of alleles in a population, Mol. Ecol. 12 (2003) 3157-3164.

[35] Rajora O.P., Genetic biodiversity impacts of silvicultural practices and phenotypic selection in white spruce, Theor. Appl. Genet. 99 (1999) 954-961.

[36] Scotti I., Paglia G., Magni F., Morgante M., Population genetics of Norway spruce (Picea abies Karst.) at regional scale: sensitivity of different mocrosatellite motif classes in detecting differentiation, Ann. For. Sci. 63 (2006) 485-491.

[37] Soto A., Lorenzo Z., Gil L., Nuclear microsatellites markers for the identification of *Quercus ilex* L. and *Quercus suber* L. hybrids, Silvae Genet. 52 (2003) 63-66.

[38] Spencer C.C., Neigel J.E., Leberg P.L., Experimental evaluation of the usefulness of microsatellite DNA for detecting demographic bottlenecks, Mol. Ecol. 9 (2000) 1517-1528.

[39] Steinkellner H., Fluch S., Turetschek E., Lexer C., Streiff R., Kremer A., Burg K., Glössl J., Identification and characterization of (GA/CT)n - microsatellite loci from *Quercus petraea*, Plant Mol. Biol. 3 (1997) 1093-1096.

[40] Stoehr M.U., El-Kassaby Y.A., Levels of genetic diversity at different stages of the domestication cycle of interior spruce in British Columbia, Theor. Appl. Genet. 94 (1997) 83-90.

[41] Streiff R., Ducousso A., Lexer C., Steinkellner H., Glössl J., Kremer A., Pollen dispersal inferred from paternity analysis in a mixed oak stand of *Quercus robur* L. and *Q. petraea* (Matt.) Liebl., Mol. Ecol. 8 (1999) 831-841.

[42] Van Osterhoout C., Hutchinson W.F., Wills D.P.M., Shipley P., MICRO-CHECKER: software for identifying and correcting genotyping errors in microsatellite data, Mol. Ecol. Notes 4 (2004) 535-538.

[43] Vencovsky R., Crossa J., Variance effective population size under mixed self and random mating with applications to genetic conservation of species, Crop Sci. 39 (1999) 1282-1294.

[44] Wang J., A pseudo-likelihood method for estimating effective population size from temporally spaced samples, Genet. Res. 78 (2001) 243-257.

[45] Weir B.S., Cockerham C.C., Estimating *F*-statistics for the analysis of population structure, Evolution 38 (1984) 1358-1370.

**Figure 1**

Location of the sampled populations of *Quercus ilex* in Sicily: seed origin stand Piano Zucchi forest (A), Piano Noce nursery (B) and Monte Palmeto plantation (C). An enclosed group of 15 holm oaks (D) is the seed-tree source of nursery and plantation. Relative position of trees and diameter to the breast height (represented as a circle size) are shown.

**Figure 2**

Distribution of maternity assignments to the 15 candidate mother trees. Individuals from the nursery and the plantation are assigned with two confidence levels 80% and 95%.

533  **Table I** Mean value of diversity indices (number of alleles $n_a$, allelic richness $A$,
534  effective number of allele $A_e$ and expected heterozygosity $H_e$) for six microsatellite
535  loci and genetic differentiation ($\theta$) per pair of populations.
536

|       | seed origin stand | nursery | P-value | seed origin stand | plantation | P-value |
|-------|-------------------|---------|---------|-------------------|------------|---------|
| $n_a$ | 10                | 7       | 0.0007  | 10                | 6          | < 0.0001 |
| $A$   | 8.795             | 6.978   | 0.0009  | 8.795             | 5.974      | 0.0001  |
| $A_e$ | 4.536             | 3.681   | 0.0098  | 4.536             | 3.062      | < 0.0001 |
| $H_e$ | 0.653             | 0.641   | 0.2474  | 0.653             | 0.518      | < 0.0001 |
| $\theta$ | 0.023          |         | 0.0028  | 0.036             |            | 0.0002  |

537

538

539  **Table II** Average non-exclusion probabilities in one candidate parent (NE-1P) and
540  candidate parent pair (NE-PP) assignations calculated by CERVUS 3.0, separately for
541  the six microsatellite loci and combined for two sets of six and five loci (without
542  QpZAG36). Loci sorted by increasing values.

543

| Loci | NE-1P | NE-PP |
| --- | --- | --- |
| QrZAG20 | 0.457 | 0.124 |
| QrZAG11 | 0.460 | 0.125 |
| QpZAG15 | 0.592 | 0.232 |
| QpZAG46 | 0.665 | 0.277 |
| QpZAG36 | 0.828 | 0.516 |
| MSQ4 | 0.995 | 0.903 |
| **Combined: 6 loci** | 0.068 | <0.001 |
| **Combined: 5 loci** | 0.082 | 0.001 |

544

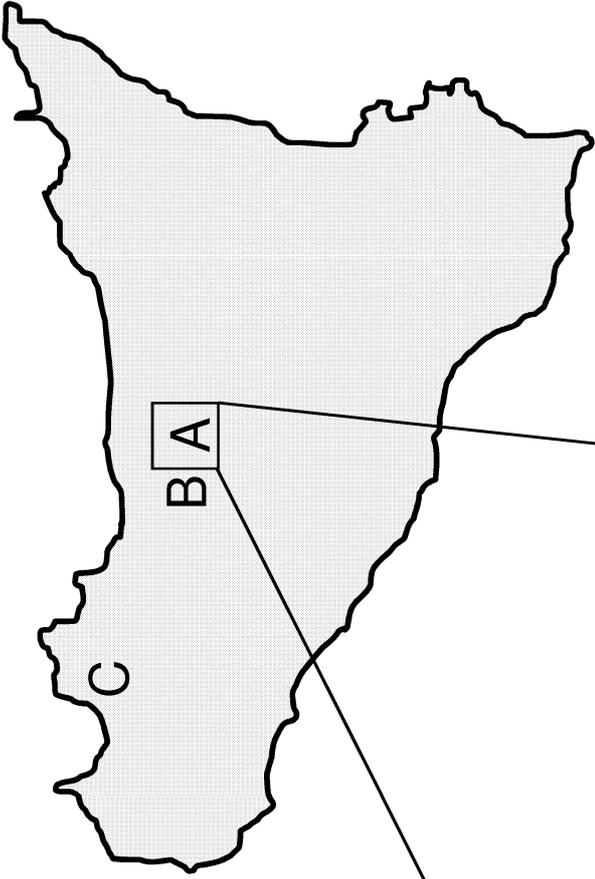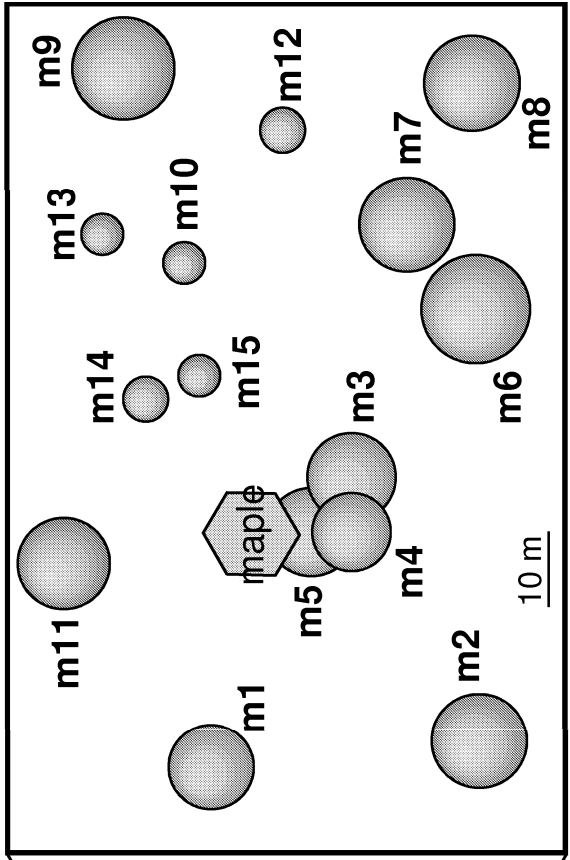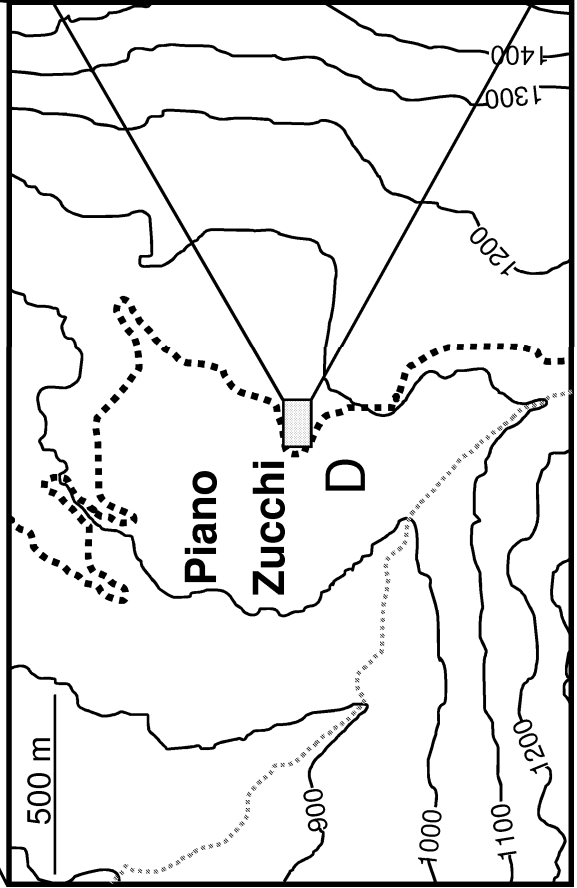

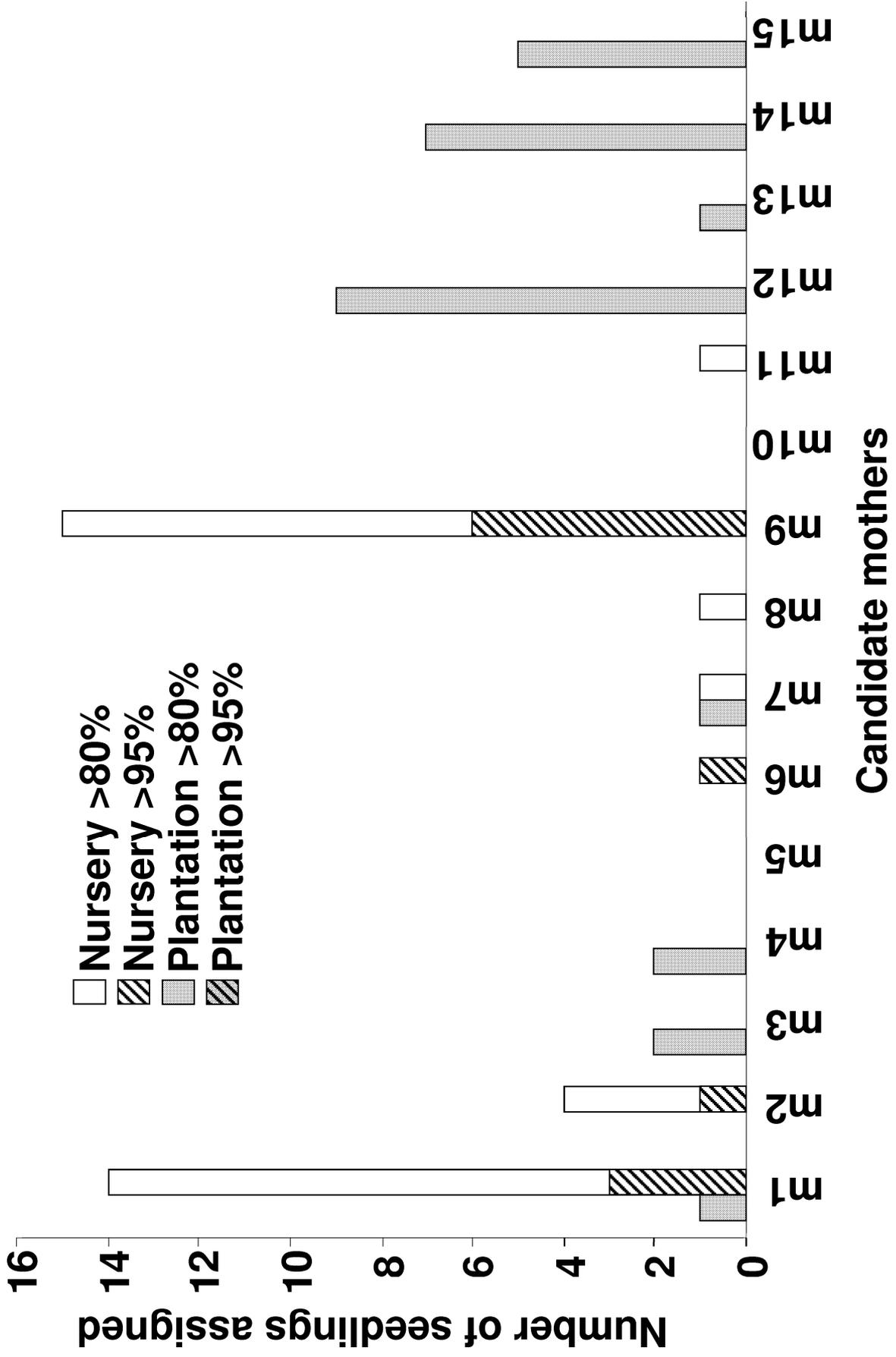